# Hybrid Learning Aided Inactive Constraints Filtering Algorithm to Enhance AC OPF Solution Time

Fouad Hasan, *Student Member, IEEE,* Amin Kargarian, *Member, IEEE,* Javad Mohammadi, *Member, IEEE*

*Abstract*—The Optimal power flow (OPF) problem contains many constraints. However, equality constraints along with a limited set of active inequality constraints encompass sufficient information to determine the problem's feasible space. In this paper, a hybrid supervised regression-classification learning-based algorithm is proposed to identify active and inactive sets of inequality constraints of AC OPF solely based on nodal power demand information. The proposed algorithm is structured using several classifiers and regression learners. The combination of classifiers with regression learners enhances the accuracy of active/inactive constraints identification procedure. The proposed algorithm modifies the OPF feasible space rather than a direct mapping of OPF results from demand. Inactive constraints are removed from the design space to construct a truncated AC OPF. This truncated optimization problem can be solved faster than the original problem with less computational resources. Numerical results on several test systems show the effectiveness of the proposed algorithm for predicting active and inactive constraints and constructing a truncated AC OPF. We have posted our code for all simulations on arxiv and have uploaded the data used in numerical studies to IEEE DataPort as an open access dataset.

## Nomenclature

| | |
|---|---|
| $g$ | Index for generators. |
| $i, j$ | Index for buses. |
| $l$ | Index for branches. |
| $k$ | Index for demand samples. |
| $n_b$ | Set of buses with nonzero demand. |
| $n_{b'}$ | Set of buses with nonzero net injection. |
| $n_g$ | Set of generators. $v, \delta$ Voltage magnitude and angle. |
| $D$ | Nodal power demand vector. |
| $F_{max}$ | Maximum branch flow. |
| $G$ | Actual generation vector. |
| $P_d, Q_d$ | Real and reactive power demand. |
| $p_{di}^L$ | Minimum value of load at bus $i$. |
| $p_{di}^U$ | Maximum value of load at bus $i$. |
| $p_g, q_g$ | Actual real and reactive power generation. |
| $S$ | Complex power. |
| $V_m$ | Voltage magnitude. |
| $\theta_i$ | Voltage angle of bus $i$. |
| $NI_P, NI_Q$ | Actual net real and reactive power injection. |
| $h_v(x)$ | Set of voltage constraints. |
| $h_l(x)$ | Set of branch flow constraints. |
| $A(\cdot)$ | Set of true active constraints. |
| $\tilde{A}(\cdot)$ | Set of predicted active constraints by classifiers. |
| $\tilde{x}$ | Predicted $x$ values by learners. |
| $\Delta_{di}$ | Maximum perturbation range for load at bus $i$. |

## I. Introduction

OPTIMAL power flow (OPF) is one of the main energy management functions that is solved every 5~15 minutes for power system scheduling and analysis [1, 2]. The size of the OPF problem depends on multiple factors, such as the number of buses and branches. Equality and inequality constraints represent characteristics of the power system and equipment. These constraints form the OPF feasible space (also known as feasible design space, feasible region, or design space).

Because of nonconvex and complex nature of AC OPF, solving this problem for large systems is computationally expensive and time-consuming. Various approaches have been proposed in the literature to reduce the computational cost of OPF. Since the majority of OPF inequality constraints are inactive in most cases, one potential approach for relieving computational costs of OPF is to identify inactive constraints and omit them from the optimization. There are a few papers for the identification of active and inactive constraints for OPF applications. Most of these papers rely on mathematical and optimization approaches to identify OPF inactive constraints. Reference [3] discusses that over 85% of branch constraints are inactive in security-constrained unit commitment (SCUC) problems. An analytical condition is developed to identify the set of inactive branch constraints for DC optimal power flow formulation. The concept of umbrella constraints is presented in [4] to describe the feasible set of DC OPF with necessary and sufficient constraints aiming at reducing the size of the problem and making it less computationally expensive. This reference presents a mathematical optimization method that finds the umbrella constraints. Authors in [5] have proposed a framework to reduce the number of security constraints in SCUC. An optimization-based bound tightening scheme is presented that solves multiple linear programs in parallel to identify redundant linear security constraints. Each linear program contains fewer constraints than the original SCUC. It is observed that roughly 99% of constraints are redundant in real-world scenarios. The proposed algorithm requires the information of network topology and upper and lower bound of nodal injection and branch flow limits. The algorithm is independent of unit commitment parameters and uncertain load values. Moreover, [6] proposes an iterative contingency search algorithm that can remove the majority of inactive transmission constraints form the SCUC problem. Linear sensitivity factors are used to find violated constraints.

This work was supported by the National Science Foundation under Grant ECCS-1944752.
F. Hasan and A. Kargarian are with the Electrical and Computer Engineering Department, Louisiana State University, Baton Rouge, LA 70803, (e-mail: fhasan1@lsu.edu, kargarian@lsu.edu).
J. Mohammadi is with the Electrical and Computer Engineering Department, Carnegie Mellon University, Pittsburgh, PA 15213, (e-mail: jmohammadi@cmu.edu).

These approaches either solve sub-optimization problems or implement iterative search techniques to find active constraints. Several of these approaches, however, might be more computationally expensive than the original optimization problem. Also, these approaches are mainly developed based on convex DC OPF, not nonconvex AC OPF. Solving AC OPF is becoming of more interest in the power system community. Hence, innovative approaches are required for active/inactive constraints identification for the AC OPF problem.

This paper presents a combined learning and model-based algorithm to speed up AC OPF solution time. In the learning phase, a hybrid supervised regression-classification based approach is proposed to identify active and inactive bus voltage and branch flow constraints of the AC OPF problem. The proposed algorithm reads nodal real and reactive power demand as inputs and predicts the set of active inequality constraints with the aim of reducing the size of OPF in a fast and efficient manner. To enhance the accuracy of constraints status identification, two regression learners are trained to project generating units' production by reading demand information, and then the outputs of these learners are used along with demand information to train two classifiers, one for voltage constraints and another for branch flow constraints. As shown in Fig. 1, the proposed algorithm constructs a truncated AC OPF with a subset of constraints containing sufficient information for forming the OPF feasible design space. This makes the proposed algorithm different than several existing methods that directly predict OPF results from demand. The solutions of the truncated and original AC OPF problems are the same while solving the truncated optimization is much faster and needs less computational resources. The simulation results show the effectiveness of the proposed algorithm for active and inactive inequality constraints classification and constructing a truncated AC OPF. **We have posted our code for all simulations on arxiv and have uploaded the data used in case studies to IEEE DataPort as an open access dataset (DOI: 10.21227/kege-qv50).**

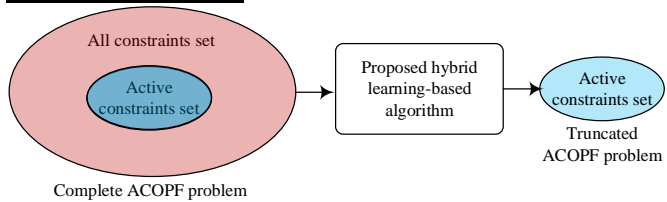

Fig. 1. The proposed active constrains filtering strategy.

The remainder of the paper is organized as follows. Section II provides a brief overview of the ongoing applications of machine learning in solving various OPF related problems. The proposed algorithm is presented in Section III. Section IV provides information about the learners used in this paper. Section V demonstrates the numerical simulation results. Section VI provides concluding remarks, and future work is discussed in Section VII.

## II. RELATED WORK

Machine learning is a collection of algorithms that enables a machine to learn from observation and analysis without any external influence [7]. The applications of machine learning to solve power systems problems [8, 9], particularly OPF, has gained a growing interest in recent years [10]. Most of the recently published papers focus on the direct projection of OPF solution by machine learning tools using demand information as inputs to learners while ignoring the knowledge of the known mathematical structure of the OPF problem [11-16]. This approach works like a black-box that read demand and estimates OPF results.

A supervised machine learning-based security-constrained OPF framework is developed in [11] that uses multi-target regression to directly map the local information and generation dispatches. This framework uses local features as inputs to machine learning models. Reference [12] provides a direct mapping of OPF results using gradient boosting regression. Demand and production cost information is used as inputs to learners that predict power and voltage of each generator. Nearest neighbor classification, which is a machine learning algorithm, is used in [13] to provide an approximate unit commitment solution for market-clearing without the need for computationally expensive unit commitment solvers. Demand and wind generation are inputs to learners, and unit commitment decisions are outputs. In [14], machine learning is applied to predict OPF results to regulate voltage and power flow in distribution grids. In [14] and [15], the proposed method implements a decentralized OPF based reactive power controller using multiple linear regression learners. This method is implemented on a system with multiple controllable distributed energy resources (DERs). In [17], support vector machine (SVM) is used to implement the Volt-VAR control scheme. Different SVM kernels like linear, polynomial, and radial basis functions are compared by the lowest sample mean squared error. The OPF formulation considers the uncertainty coming from renewable energy sources and load. In [18], the authors have extended their work and presented a machine learning-based method to predict optimal settings of a centralized controller based on historical data. While only inverter-based DER reactive power controller is considered in [17], active power curtailment, controllable load shifting, and battery storage are taken into consideration in [18]. Reference [19] has proposed a machine learning-based approach for transient stability constrained OPF based on critical clearing time constraints. Multilayer feedforward neural network is used to compute the critical clearing time of the formulated OPF problem. Deep learning is used in [20] to predict OPF results. This approach is applicable if at a state of a system, the information of prior states of the system is available to learners.

Such direct estimations, however, do not precisely match with actual solutions. While a trained learner might provide good estimations for many loading conditions, it might not provide accurate enough solutions for many other demand scenarios. An immense training dataset might be required to reach an acceptable level of accuracy for learners. Even if the accuracy of direct OPF solution estimation algorithms is high, a small mismatch between projected and actual solutions results may yield a suboptimal or infeasible solution for the nonlinear, nonconvex AC OPF problem. This makes operators reluctant to deploy them for real power systems operation. One may use a combined learning and model-based approach to reduces the possibility of suboptimality and infeasibility of OPF results. The benefits of learning based warm start to solve AC OPF are presented in [21]. Instead of solving OPF directly with machine learning, the demand information is used as inputs to learners to estimate the OPF solution. This solution is used as a starting point to solve the OPF problem with the complete set of constraints. Although having a warm start enhances solution speed, this method does not reduce the size of the OPF problem that has a significant impact on the computational complexity of AC OPF.

An idea recently presented in a few papers is to use machine learning to predict inactive constraints rather than using machine learning tools as black boxes to directly predict OPF results [22-28]. In [22, 23], an approach is presented to learn the mapping from uncertainty realization to the optimal solution. This approach avoids directly mapping the input to the optimal solution and instead uses the sets of active constraints at optimality as the mapping output. This approach does not require the continuous mapping of inputs and outputs of a complex system rather simplifies the learning task and utilizes the known mathematical model of the system. Reference [24] presents another approach to learn the set of active constraints at the optimal point using classification algorithms. A neural network classifier is used for learning the active sets. This paper deals with DC OPF and uses only classification learners. The authors of [25] have presented a learning based method to predict umbrella constraints for an OPF problem. The umbrella constraints are necessary and sufficient constraints to cover the OPF feasible solution. References [26] and [27] present a learning-based chance-constrained approach to remove constraints with zero probability events from the AC OPF formulation for distribution networks. With the help of statistical learning, the proposed framework reduces the computationally demanding joint chance constraints into a series of single chance constraints. Reference [28], which serves as a modified version of the algorithm presented in [6], uses machine learning to predict redundant transmission constraints, warm start, and an affine subspace that contains the optimal solution of SCUC.

While these approaches are promising, they mainly focus on DC OPF. More sophisticated yet efficient algorithms are needed to detect inactive constraints of the AC OPF problem. These papers use the demand information and train a classifier(s) to identify the status of constraints. A combination of regression and classification learners can enhance the accuracy of the constraint identification process. In this paper, we investigate developing a hybrid regression-classification-based algorithm for identifying the status of constraints of the AC OPF problem.

## III. Hybrid Regression-Classification Algorithm for Inactive Constraints Identification

### A. Classical AC OPF Formulation

The considered AC OPF problem, presented by (1a)-(1i), is adopted from [29]. The objective function is to minimize generation costs. Nodal power balance constraints are given by (1b) and (1c). Constraints (1d) and (1e) enforce branch flow limits at two ending terminals of a branch. The upper and lower bounds of generating units are imposed by (1f) and (1g). Inequalities (1h) and (1i) are bus voltage magnitude and angle limits.

$$\min f(p) = \sum_g a_g \cdot p_g^2 + b_g \cdot p_g + c_g \qquad (1a)$$

s.t.

$$g_p(\theta, V_m, p_g) = P_{bus}(\theta, V_m) + P_d - p_g = 0 \qquad (1b)$$

$$g_q(\theta, V_m, q_g) = Q_{bus}(\theta, V_m) + Q_d - q_g = 0 \qquad (1c)$$

$$h_{ls}(\theta, V_m) = |F_{ls}(\theta, V_m)| - F_{max} \leq 0 \qquad (1d)$$

$$h_{lr}(\theta, V_m) = |F_{lr}(\theta, V_m)| - F_{max} \leq 0 \qquad (1e)$$

$$p_g^{min} \leq p_g \leq p_g^{max} \qquad \forall g \qquad (1f)$$

$$q_g^{min} \leq q_g \leq q_g^{max} \qquad \forall g \qquad (1g)$$

$$V_i^{min} \leq V_i \leq V_i^{max} \qquad \forall i \qquad (1h)$$

$$\theta_i^{ref} \leq \theta_i \leq \theta_i^{ref} \qquad \forall i \qquad (1i)$$

### B. Constraints Status Identification

To construct a truncated OPF, inactive inequality constraints should be detected and omitted from the optimization problem. Detecting active constraints (constraints that have reached their specified minimum or maximum limits) and inactive constraints is a binary classification problem. Without loss of generality, we focus on identifying the status of bus voltage magnitude and branch flow constraints. These two sets of inequalities have high impacts on OPF computation cost. The total number of voltage magnitude and branch flow constraints is higher than that of other OPF inequalities, e.g., generators upper and lower bounds, while the majority of these two sets of constraints are inactive under various loading conditions. This is not a valid argument for generators limits as constraints of many of these controllable devices might be active under several loading conditions.

The objective is to predict constraints status using only nodal demand values. For brevity, we represent branch flow constraints (1d) and (1e) and voltage magnitude constraints (1h) in compact forms as follows:

$$h_l(x) := \{h_{ls}(\theta, V_m); h_{lr}(\theta, V_m)\} \qquad (2)$$

$$h_v(x) := \{V_i^{min} - V_i \leq 0; V_i - V_i^{max} \leq 0\} \qquad (3)$$

Since the bus voltage and branch flow constraints are inherently different, we train two separate classifiers with one for bus voltage constraints and another one for branch flow constraints.

*Dataset Preparation:* Before solving OPF, demand information is available. We define the following demand vector $D$ as the input for learners.

$$P_d = [p_{d1}, p_{d2}, \ldots, p_{dn}]^T \qquad \forall n \in n_b \qquad (4a)$$

$$Q_d = [q_{d1}, q_{d2}, \ldots, q_{dn}]^T \qquad \forall n \in n_b \qquad (4b)$$

$$D = \begin{bmatrix} P_d \\ Q_d \end{bmatrix} \qquad (4c)$$

To cover possible loading situations that may occur during system operation in the training phase, we generate a set of demand scenarios as follows:

$$p_{di}^k = [p_{di}^l + \eta_p(k) \cdot \Delta_{di}] \qquad \forall k \qquad (5a)$$

$$q_{di}^k = [q_{di}^L + \eta_q(k) \cdot \Delta_{di}] \quad \forall k \quad (5b)$$

$$\Delta_{di} = p_{di}^U - p_{di}^L \quad (5c)$$

where $\eta_p(\cdot)$ and $\eta_q(\cdot)$ follows a uniform distribution between 0 and 1. The perturbation range $\Delta_{di}$ depends on the possible minimum ($P_d^L$) and maximum values ($P_d^U$) of the total system load and might not be the same for all load points. For each demand scenario, OPF is solved and active and inactive bus voltage ($A(h_v(x))$) and branch flow constraints ($A(h_l(x))$) are identified and stored for training. Several demand scenarios may result in infeasible OPF solutions. These scenarios are abandoned during the training dataset preparation.

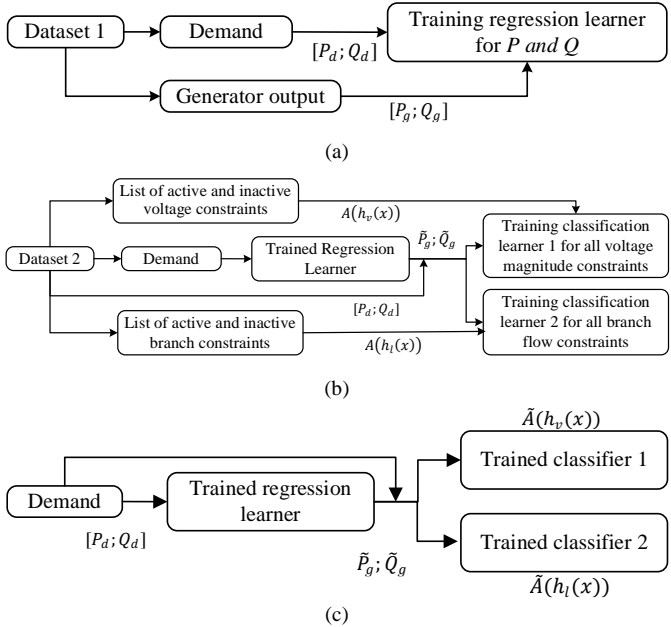

Fig. 2. Block diagram of a) regressor training procedure, b) classifiers training procedure, and b) the utilization of trained learners.

*Proposed Training Structure:* The status of voltage and branch flow constraints not only depends on demand values but also is affected by generating units' production. Using only demand data for classification purposes might degrade the accuracy of trained classifiers. However, only demand information is known before solving OPF. Generation values will be known after solving the OPF problem. To augment the classifiers' input features, we add a regression step, as shown in Fig. 2a, to the proposed algorithm and predict generation values from demand. A regressor is trained whose input and target vectors are $D$ and $[P_g, Q_g]$. The input vector to the regressor is which contains the nodal real and reactive power demand. The trained regressor will provide predicted real ($\tilde{P}_g$) and reactive power generation ($\tilde{Q}_g$) for a given demand vector $D$. By using one regressor for predicting $\tilde{P}_g$ and $\tilde{Q}_g$, the learner captures information of real and reactive powers and better understands generator dynamics and interaction between real and reactive powers. The input of the regression learner is nodal demand, and its output is the predicted power generated by each unit.

$$\tilde{P}_g = [\tilde{p}_{g1}, \tilde{p}_{g2}, \cdots, \tilde{p}_{gn}]^T \quad \forall n \in n_g \quad (6a)$$

$$\tilde{Q}_g = [\tilde{q}_{g1}, \tilde{q}_{g2}, \cdots, \tilde{q}_{gn}]^T \quad \forall n \in n_g \quad (6b)$$

$$\tilde{G} = \begin{bmatrix} \tilde{P}_g \\ \tilde{Q}_g \end{bmatrix} \quad \forall n \in n_g \quad (6c)$$

Vector $D$ and predicted nodal power generation $\tilde{G}$ are used to form a net nodal power injection vector ($\widetilde{NI}$). Instead of complete sets of nodal demand and generation vectors, the subset of buses with nonzero demand/generation is used to create input and target vectors for classifiers and regression learners. Several buses may have neither load nor generation. Having these buses in the net nodal power injection vector provides no meaningful information for learners as their corresponding entries in the net injection vector are always zero with fixed locations.

$$\widetilde{NI}_P = [\tilde{p}_{g1} - p_{d1}, \tilde{p}_{g2} - p_{d2}, \ldots, \tilde{p}_{gn} - p_{dn}]^T \quad \forall n \in n_{b'} \quad (7a)$$

$$\widetilde{NI}_Q = [\tilde{q}_{g1} - q_{d1}, \tilde{q}_{g2} - q_{d2}, \ldots, \tilde{q}_{gn} - q_{dn}]^T \quad \forall n \in n_{b'} \quad (7b)$$

$$\widetilde{NI} = \begin{bmatrix} \widetilde{NI}_P \\ \widetilde{NI}_Q \end{bmatrix} \quad \forall n \in n_{b'} \quad (7c)$$

A possible approach is to train one classifier for each constraint. But the training task might be intractable for large systems. We train two classifiers for each system, one classifier (classifier 1 in Fig. 2a) for all sets of voltage constraints and another classifier (classifier 2 in Fig. 2a) for all sets of branch flow constraints. The input vector to the voltage and branch flow constraints classifiers is $\widetilde{NI}$, and their outputs are $\tilde{A}(h_v(x))$ and $\tilde{A}(h_l(x))$. The pseudocode to train the learners is represented in Algorithm I.

Note that different datasets are used to train the regressor and classifiers. The output of regressor along with demand is used in forming the input ($\widetilde{NI}$) to the classifiers during training. Thus, using the same dataset to train the regressor and classifiers may give a false indication of high accuracy. Dataset 1 is used to train the regressor. Then, dataset 2 is fed into the trained regressor, and corresponding predicted generations are stored. The predicted generations and demand from dataset 2 are used to form a net injection vector that is used as the input to train the classifiers.

**Algorithm I Proposed Training Architecture**

1. **Dataset 1**: Generate demand scenarios $D = \begin{bmatrix} P_d \\ Q_d \end{bmatrix}$ using (5a) and (5b)
2. Solve OPF for each scenario and drop scenarios with infeasible solution
3. Collect generation data and form $G = \begin{bmatrix} P_g \\ Q_g \end{bmatrix}$
4. Feed $D$ as input and $G = \begin{bmatrix} P_g \\ Q_g \end{bmatrix}$ as targets to regression learner and train the learner
5. **Dataset 2**: Generate a set of new demand scenarios using (5a) and (5b)
6. Do Step 2 for dataset 2
7. Check (2) and (3) and categorize active ($A(h_v(x))$ and $A(h_l(x))$) and inactive voltage and branch flow constraints for dataset 2
8. Use trained regressor in Step 4 and determine predicted generation $\tilde{G} = \begin{bmatrix} \tilde{P}_g \\ \tilde{Q}_g \end{bmatrix}$ for demand scenarios in dataset 2
9. Form $\widetilde{NI} = \begin{bmatrix} \widetilde{NI}_P \\ \widetilde{NI}_Q \end{bmatrix} = \begin{bmatrix} \tilde{P}_g - P_d \\ \tilde{Q}_g - Q_d \end{bmatrix}$ using $D$ and $\tilde{G}$ of Steps 5 and 8
8. Feed $\widetilde{NI}$ as input and $A(h_v(x))$ as targets to classification learner 1 and train this classifier
8. Feed $\widetilde{NI}$ as input and $A(h_l(x))$ as targets to classification learner 2 and train this classifier

*Utilization Procedure:* The utilization procedure of the proposed algorithm is demonstrated in Fig. 2c. For a given demand, $\tilde{P}_g$ and $\tilde{Q}_g$ are determined by the trained regression learner. The given $D$ and the predicted $\tilde{G}$ will be used to form vector $\widetilde{NI}$ that is the input of trained classifiers one and two. The output of classifier one is active bus voltage constraints ($\tilde{A}(h_v(x))$), and the second classifier predicts active branch flow constraints ($\tilde{A}(h_l(x))$). $\tilde{A}(h_v(x))$ and $\tilde{A}(h_l(x))$ will be used to construct a truncated optimization design space and consequently a truncated OPF problem as:

$$\min \sum_g a_g \cdot p_g^2 + b_g \cdot p_g + c_g \quad (8a)$$

s.t.

$$\tilde{A}(h_v(x)) \leq 0 \quad (8b)$$
$$\tilde{A}(h_l(x)) \leq 0 \quad (8c)$$
$$x \in \chi$$

where $\chi$ represents all other constraints except for bus voltage magnitude and branch flow constraints. If all active constraints are predicted correctly, the feasible space of the truncated OPF problem is the same as that of the original OPF while its size is much smaller than the original optimization problem. The pseudocode to utilize the proposed regression-classification technique to form the truncated OPF is as follows.

**Proposed Utilization Algorithm**

1. For a given demand vector $D = \begin{bmatrix} P_d \\ Q_d \end{bmatrix}$, run the trained regressor to determine $\tilde{G} = \begin{bmatrix} \tilde{P}_g \\ \tilde{Q}_g \end{bmatrix}$
4. Form $\widetilde{NI} = \begin{bmatrix} \widetilde{NI}_P \\ \widetilde{NI}_Q \end{bmatrix} = \begin{bmatrix} \tilde{P}_g - P_d \\ \tilde{Q}_g - Q_d \end{bmatrix}$ using $D$ and $\tilde{G}$
5. Use $\widetilde{NI}$ as the input to the trained classifiers 1 and 2
6. Identify active voltage $\tilde{A}(h_v(x))$ and branch flow $\tilde{A}(h_l(x))$ constraints from the classifiers' outputs
7. Construct the truncated AC OPF problem
8. Minimize the objective function (8a) subject to (8b), (8c), and $\chi$

One may use the regression learner of Fig. 2 to predict $\tilde{P}_g$ and $\tilde{Q}_g$ and then formulate and solve a modified AC power flow instead of a truncated AC OPF. Although solving AC power flow is easier than solving the truncated AC OPF, even a slight error in $\tilde{P}_g$ and $\tilde{Q}_g$ might make AC power flow results suboptimal and, more importantly, endanger power flow feasibility.

## IV. SELECTING LEARNING APPROACH AND ALGORITHM

Supervised learning approaches must be selected to train classifiers and regression learners in Fig. 2 as the training datasets are labeled. A wide range of supervised machine learning approaches can be used. We have examined support vector machine (SVM) with quadratic and Gaussian functions, Gaussian process regression with exponential and quadratic kernels, and ensemble learning with bagging and boosting methods for regression learners. We have also examined SVM with coarse quadratic and Gaussian functions, the k-nearest neighbor with coarse and weighed techniques, discriminant analysis with linear and quadratic functions, and Naïve Bayes for classification learners. We have observed that while the performance of these approaches is suitable for small power systems, their performance degrades by increasing the size of the system. We have also observed cases in which these learners failed to map a function between the input and output AC OPF training datasets.

We have tested neural networks (NNs) on power systems with different sizes and have observed that NN shows good performance for mapping power generation from demand and classifying voltage and branch flow constraints. Hence, we have selected NN for regression and constraints classification. Using activation functions, NN can capture the nonlinearity and complexity of problems, such as AC OPF, effectively. We have used fully connected NN regression and classification learners with mini batch gradient descent. We use regression learners with Rectified linear units (ReLU) in hidden layers and linear activation functions in the output layer, whereas we use the classification learners with the sigmoid function in hidden layers and the softmax function in the output layer. In the case of linear activation function, the output is proportional to the provided input ($X(z) = (mZ)$) whereas, based on the input, the sigmoid activation function provides the output between 0 and 1 ($X(z) = \frac{1}{1+e^{-z}}$). Softmax is a type of sigmoid function that is used for classification ($X(z) = \frac{e^{Z_k}}{\sum e^{Z_k}}$) in the final layer of neural network. The derivative of the activation function is used in the error backpropagation algorithm, which is a process to optimize the weights of each neuron. A mean squared error (MSE) loss function is used for regression problem.

$$MSE = \frac{\sum_{k=1}^{K}(X^k - \tilde{X}^k)^2}{K} \quad (9)$$

Both real and reactive power are combinedly used to update the weights ($\frac{\sum_{k=1}^{K}[(P_g^K - \tilde{P}_g^K)^2 + (Q_g^K - \tilde{Q}_g^K)]}{K}$). And a binary cross-entropy (BCE) loss function is used for classification problems.

$$BCE = -\frac{\sum_{k=1}^{K}[X^k \log(\tilde{X}^k) + (1-X^k)\log(1-\tilde{X}^k)]}{K} \quad (10)$$

where $X^k$ and $\tilde{X}^k$ are true values and predicted values. Adam optimizer is used to train regressors and classifiers. We have tested various architectures with different numbers of layers, epochs, and batch sizes. Figure 3 illustrates the results obtained by regressors with different numbers of hidden layers. Merely increasing the number of layers does not improve the prediction accuracy for all systems. We have selected one hidden layer for regressors. Table I depicts the learners' architecture and hyperparameters used in this paper. Note that although we have obtained promising results with these simple architectures, one can use more complex architectures to obtain even better results.

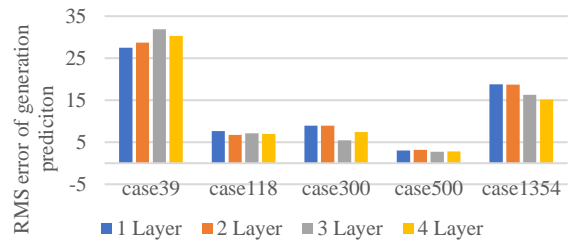

Fig. 3. RMSE comparison of different NN regression architectures.

TABLE I
ARCHITECTURE OF TRAINED NNS

| Learner | Training parameters | Activation function | Loss function | Optimizer |
|---|---|---|---|---|
| Regressor | hidden layer=1 neuron=256 epochs=1000, batch size=100 validation split=20% | ReLU & Linear | MSE | Adam |
| Classifier | hidden layer=1 neuron=256 epochs=1000, batch size=100 validation split=20% | Sigmoid & SoftMax | BCE | Adam |

## V. NUMERICAL RESULTS

The effectiveness of the proposed algorithm for detecting active and inactive constraints is tested on several small, medium, and large systems. Test systems are adopted from the standard PGLib-OPF benchmark library [30]. MATPOWER interior point solver is used to solve OPF [29]. Python (v3.7.3) based Keras framework (v2.3.1) is used with TensorFlow backend during the learning phase. Simulations are carried out on a personal computer with a 3.70 GHz Intel(R) Xeon(R) CPU, eight cores, and 16 GB of RAM. **We have posted our code for all simulations on arxiv and have uploaded the data used in numerical studies to IEEE DataPort an open access dataset (DOI: 10.21227/kege-qv50).**

### A. Average Number of Active and Inactive Constraints

We have analyzed the number of active and inactive constraints for several test systems. Table II shows the number of voltage and branch constraints for the original OPF and truncated OPF problems. The second column shows the total number of voltage and branch constraints, and the third column depicts the average number of active voltage and branch constraints under various loading conditions. For the 39-bus system, for instance, the total number of voltage and branch constraints are 78 and 92, respectively, out of which, on average, five voltage constraints and two branch flow constraints are active. It is observed that larger systems have a higher percentage of inactive constraints. This shows the potential advantage of detecting active constraints to construct a truncated OPF problem instead of the original OPF. For the 39-bus and 118-bus systems, for instance, the number of constraints of the truncated OPF problem is on average 55% (including all equality and inequality constraints of (1)) less than that of the original OPF.

TABLE II
NUMBER OF TOTAL CONSTRAINTS AND ACTIVE CONSTRAINTS FOR SEVERAL TEST SYSTEMS

| System | Original OPF (Voltage, Branch flow) | Truncated OPF (Active Voltage, Branch flow) | Inactive, Active |
|---|---|---|---|
| case39_epri | 78, 92 | 5, 2 | 96%, 4% |
| case118_ieee | 236, 372 | 12, 2 | 99%, 1% |
| case300_ieee | 600, 822 | 32, 3 | 97.5%, 2.5% |
| case500_tamu | 1000, 1192 | 16, 4 | 99%, 1% |
| case1354_pegase | 2708, 3982 | 50, 20 | 99%, 1% |

### B. Inactive Constraints Identification by Proposed Hybrid Algorithm

*Training:* Nodal power demand is varied using uniform random distribution to generate possible demand scenarios over a long operation horizon. Table III shows the load perturbation range ($\Delta_d$) as compared to MATPOWER baseload. The range is obtained by monotonically decreasing and increasing the base case load until the simulation fails to converge. This range is narrower for the larger systems. OPF is solved for each demand scenario. One regression learner is trained for each system. As shown in Table IV, the length of the regression learners' output is equal to twice the number of generators, whereas the length of branch flow (voltage) constraints classifier's output is equal to the number of branches (buses). Active voltage and branch flow constraints are labeled as '1', and inactive constraints are labeled as '0' during the preparation of datasets. Two classifiers are trained for each test system. Note, for ease of replication of simulations and to show the performance of the proposed algorithm with simple machine learning architectures, we have used the same architecture for all learners.

TABLE III
SYSTEM PARAMETERS AND RANGE OF VARIATION OF LOAD

| System | NB/NL/NG | $\Delta_d$ | No of Scenario | | |
|---|---|---|---|---|---|
| | | | Regressor (Dataset1) | Classifier (Dataset2) | Testing |
| case39 | 39/46/10 | 70% to 130% | 2000 | 2000 | 882 |
| case118 | 118/186/54 | 70% to 130% | 2000 | 2000 | 2000 |
| case300 | 300/411/69 | 92% to 104% | 2000 | 2000 | 1641 |
| case500 | 500/597/90 | 70% to 109% | 2000 | 2000 | 3000 |
| case1354 | 1354/1991/260 | 70% to 110% | 1500 | 1500 | 1200 |

\* NB/NL/NG stands for number of Node, Branch and Generator respectively

TABLE IV
INPUT AND OUTPUT LENGTHS OF LEARNERS

| System | Regression learner | | Classifiers | | |
|---|---|---|---|---|---|
| | D | $P_g; Q_g$ | NI | $h_v$ | $h_l$ |
| case39_epri | 42 | 10*2 | 78 | 39 | 46 |
| case118_ieee | 189 | 54*2 | 236 | 118 | 186 |
| case300_ieee | 374 | 69*2 | 600 | 300 | 411 |
| case500_tamu | 400 | 90*2 | 1000 | 500 | 597 |
| case1354_pegase | 1332 | 260*2 | 2708 | 1354 | 1991 |

*Testing:* For each studied system, the size of the training and test datasets is provided in Table III. For each scenario, the original OPF problem is solved to determine the actual active/inactive status of constraints, and the proposed hybrid algorithm is also applied to predict active/inactive status of constraints. Four primary indices are introduced to interpret predicted results and analyze the accuracy of the proposed algorithm.

- *True positives* (TP) are cases in which a constraint is predicted to be ACTIVE and its actual status is also ACTIVE.
- *True negatives* (TN) are cases in which the prediction is INACTIVE and the actual output is INACTIVE.
- *False positives* (FP) are cases in which the prediction is ACTIVE but the actual output is INACTIVE (type I error).

TABLE V
PREDICTION ACCURACY MEASUREMENTS OF THE PROPOSED ALGORITHM FOR VOLTAGE CONSTRAINTS CLASSIFICATION

| systems | FN | FP | TN | TP | NPV | PPV | TPR | TNR | Misclassification | Accuracy |
|---|---|---|---|---|---|---|---|---|---|---|
| case39_epri | 0.06% | 5.3% | 87.5% | 7.1% | 99.9% | 57.1% | 99.5% | 94.3% | 5.7% | 94.6% |
| case118_ieee | 0.003% | 0.53% | 97.1% | 2.6% | 100% | 90% | 100% | 99.8% | 0.20% | 99.8% |
| case300_ieee | 0.015% | 2.13% | 92.5% | 5.4% | 99.9% | 71.2% | 99.7% | 97.7% | 2.2% | 97.8% |
| case500_tamu | 0.004% | 0.9% | 97.5% | 1.6% | 99.9% | 63.3% | 99.7% | 99.1% | 0.9% | 99.1% |
| case1354_pegase | 0.007% | 0.91% | 98.2% | 0.85% | 99.9% | 48.1% | 99.1% | 99.07% | 0.9% | 99.1% |

TABLE VI
PREDICTION ACCURACY MEASUREMENTS OF THE PROPOSED ALGORITHM FOR BRANCH CONSTRAINTS CLASSIFICATION

| systems | FN | FP | TN | TP | NPV | PPV | TPR | TNR | Misclassification | Accuracy |
|---|---|---|---|---|---|---|---|---|---|---|
| case39_epri | 0% | 0.96% | 98.5% | 0.54% | 100% | 34.8% | 100% | 99% | 0.96% | 99.4% |
| case118_ieee | 0% | 0.01% | 99.2% | 0.54% | 100% | 67% | 100% | 99.7% | 0.3% | 99.7% |
| case300_ieee | 0% | 0.01% | 99.5% | 0.48% | 100% | 97.9% | 100% | 99.9% | 0.01% | 99.9% |
| case500_tamu | 0% | 0.73% | 98.9% | 0.35% | 100% | 32.5% | 100% | 99.3% | 0.7% | 99.3% |
| case1354_pegase | 0% | 0.11% | 99.5% | 0.36% | 100% | 75.9% | 100% | 99.88% | 0.12% | 99.88% |

- *False negatives* (FN) are cases in which the prediction is INACTIVE but the actual output is ACTIVE (type II error).

In addition, we use the following statistical metrics to analyze the quality of the truncated OPF in detail.

$$\text{Accuracy} = \frac{TP+TN}{TP+TN+FP+FN} \quad (11a)$$

$$\text{Misclassification} = \frac{FP+FN}{TP+TN+FP+FN} \quad (11b)$$

$$\text{True Positive Rate (TPR)} = \frac{TP}{TP+FN} \quad (11c)$$

$$\text{False Negative Rate (FNR)} = 1 - TPR \quad (11d)$$

$$\text{True Negative Rate (TNR)} = \text{Specificity} \frac{TN}{TN+FP} \quad (11e)$$

$$\text{False Positive Rate (FPR)} = 1-TNR \quad (11f)$$

$$\text{Positive Predictive Value (PPV)} = \frac{TP}{FP+TP} \quad (11g)$$

$$\text{False Discovery Rate (FDR)} = 1-PPV \quad (11h)$$

$$\text{Negative Predictive Value, NPV} = \frac{TN}{FN+TN} \quad (11i)$$

$$\text{False Omission Rate (FOR)} = 1-NPV \quad (11j)$$

Tables V and VI show these indices for several test systems. We have selected the Pegase 1354-bus test system and constructed a confusion matrix shown in Fig. 4. Each of the test scenarios contains 2708 upper/lower bus voltage magnitude constraints and 3982 sending/receiving branch flow limits. Hence, for 1200 test scenarios, the actual and predicted status of 2708×1200 voltage and 3982×1200 branch flow constraints are observed to calculate the indices shown in Fig. 4. Green blocks in the first column of Figs. 4a and 4b show that 98.2% of bus voltage constraints and 99.5% of branch constraints are true negatives, which means they are correctly predicted to be inactive. Green blocks in the second column depict that 0.85% and 0.36% of voltage and branch constraints are true positives, which means they are correctly predicted to be active. As shown in orange blocks in the second column, 0.91% of voltage constraints and 0.11% of branch constraints are misclassified to be active. This is the type I error (false positives) that means a few actual inactive constraints are predicted to be active and hence are included in the truncated OPF. This is not critical as these few constraints do not change the truncated feasible space (i.e., do not change the OPF solution) and have no considerable impact on the computational burden of the truncated OPF. The undesirable error is the type II error (false negative) that means actual active constraints are predicted to be inactive. As shown in orange blocks in the third column of the confusion matrices, the type II error is very close to zero percent. TPR for voltage and branch flow constraints is 99.1% and 100% that shows that roughly all of actual active constraints are predicted to be active. TNR pertaining to voltage and branch constraints is 99.07% and 99.88%, respectively, which shows that the percentage of actual inactive constraints that are predicted to be inactive. NPV for both voltage and branch flow constraints is 100% showing that all of predicted inactive constraints are truly inactive. In a nutshell, the accuracy indices for both types of constraints are more than 99%, and the misclassification indices are less than 1%. The misclassified constraints are mainly FP that means no important information is lost from the feasible space of the truncated OPF. Therefore, the solution of the constructed truncated OPF will be similar to the solution of the complete OPF formulation.

|  | Actual inactive | Actual active |  |
|---|---|---|---|
| Predicted inactive | 1566943<br>98.2%<br>True negative | 257<br>0.007%<br>False negative | NPV=100%<br>FOR=0.0% |
| Predicted active | 29886<br>0.91%<br>False positive | 27714<br>0.85%<br>True positive | PPV=48.1%<br>FDR=51.9% |
|  | TNR=99.07%<br>FPR=0.93% | TPR=99.1%<br>FNR=0.9% | Accuracy = 99.1%<br>Misclassification =0.9 % |

(a)

|  | Actual inactive | Actual active |  |
|---|---|---|---|
| Predicted inactive | 2366400<br>99.5%<br>True negative | 0<br>0%<br>False negative | NPV=100%<br>FOR=0.0% |
| Predicted active | 5486<br>0.11%<br>False positive | 17314<br>0.36%<br>True positive | PPV=75.9%<br>FDR=24.1% |
|  | TNR=99.88%<br>FPR=0.12% | TPR=100%<br>FNR=0% | Accuracy=99.88%<br>Misclassification =0.12% |

(b)

Fig. 4. Confusion matrices for the Pegase 1354-bus system a) voltage constraints and b) branch flow constraints.

Tables V and VI show that the FN index for all cases is negligible. We have observed a few misclassified voltage constraints. A detailed analysis reveals that these constraints are not heavily binding and have a negligible impact on the truncated feasible space. That is, including or omitting these constraints from the truncated OPF changes the optimal solution very slightly. Although no FN misclassification is observed for the majority of the studied cases, it is not guaranteed that the solution of truncated OPF always matches that of the original OPF. In such cases with nonzero FN, the solution of truncated OPF might be infeasible for the original OPF. One can apply an iterative constraints inclusion technique in addition to the predicted constraints to ensure the feasibility of the solution. Another alternative is to use the solution of truncated OPF as a warm start for the original OPF.

We use the average optimality gap as an index to measure how close are the solutions of the truncated and original OPFs.

$$\text{optimality gap\%} = \frac{|f^{T-OPF} - f^{OPF}|}{f^{OPF}} \times 100 \quad (12)$$

The values of this index, reported in Table VII, show that the solution of the truncated OPF (T-OPF) is very close to that of the original OPF.

TABLE VII
OPTIMALITY GAP OF TRUNCATED OPF

| System | Optimality gap |
|---|---|
| case39_epri | 4e-06% |
| case118_ieee | 3e-07% |
| case300_ieee | 5.9e-05 |
| case500_tamu | 7.5e-07 % |
| case1354_pegase | 3e-05% |

Table VIII shows the number of iterations of the interior point method and computation time. The time-saving values are in comparison with the original OPF. For each system, 50 loading conditions are considered, and OPF is solved ten times under each loading condition. The reported values in Table VIII are the average runtime and the average number of iterations obtained from the test scenarios. The number of iterations does not reduce significantly, but since the number of functions evaluations per iteration reduces by omitting inactive constraints, the solution time per iteration, and hence the total time reduces. The average time of each iteration can be calculated by dividing the total time to the number of iterations. For the IEEE 118-bus system, for instance, the average time of each iteration of the interior point method decreases from 6.7ms for the original OPF to 5ms for the truncated OPF, and the total time saving is 30%.

In summary, Tables VII and VIII show the promising advantage of the proposed algorithm for reducing the computation time of the AC OPF problem while providing a very high accurate solution.

TABLE VIII
ITERATION NUMBERS AND TIME-SAVING

| Systems | Number of iterations | | Time saving |
|---|---|---|---|
| | Original OPF | Truncated OPF | |
| case39_epri | 16 | 14 | 33% |
| case118_ieee | 15 | 14 | 30% |
| case300_ieee | 35 | 25 | 38% |
| case500_tamu | 25 | 18 | 35% |
| case1354_pegase | 42 | 38 | 32% |

*C. Comparison Among Three Possible Approaches*

We have tested the following three possible machine learning-based approaches to solve OPF and compared their results.

- Direct mapping of demand to optimal generation (black-box)
- Using solely demand as input to classification learners trained to identify constraints status (D to truncated OPF)
- Proposed algorithm (D and G to truncated OPF)

Mean absolute errors (MAE) of the black-box approach are not acceptable (see Fig 3). A comparison between the second and third approaches for 300-bus, 500-bus, and 1354-bus systems is represented in Table IX. Although one may obtain better results than those reported in these tables with changing the learners' architecture, to have a fair comparison, we have used the same structure (one hidden layer with 256 neurons) and hyperparameters as listed in Table I. Also, the same datasets are used for training and testing the learners. Although the demand (D) to truncated OPF approach has a lower number of FPs, the proposed approach has a higher accuracy in terms of FNs that is a crucial misclassification index for the OPF application.

TABLE IX
COMPARISON OF THREE APPROACHES

| Systems | Learner | D to truncated OPF | Proposed algorithm |
|---|---|---|---|
| Case300 | $\tilde{A}(h_v(x))$ | FN=156, FP=21062 | FN=28, FP=26778 |
| | $\tilde{A}(h_l(x))$ | FN=0, FP=133 | FN=0, FP=133 |
| Case500 | $\tilde{A}(h_v(x))$ | FN=120, FP=21062 | FN=27, FP=40057 |
| | $\tilde{A}(h_l(x))$ | FN=0, FP=12665 | FN=0, FP=12665 |
| Case1354 | $\tilde{A}(h_v(x))$ | FN=287, FP=28716 | FN=51, FP=47048 |
| | $\tilde{A}(h_l(x))$ | FN=0, FP=5486 | FN=0, FP=6668 |

## VI. CONCLUSION

In this paper, a hybrid regression-classification algorithm is proposed to identify the active and inactive sets of voltage and branch flow constraints for OPF before solving the optimization problem. It is observed that the majority of voltage and branch flow constraints are inactive, even if the system load changes, and have no impact on the OPF solution. The proposed learning algorithm identifies these inactive inequality constraints and creates a truncated OPF problem whose feasible design space includes sufficient information to encompass the optimal solution of the original complete OPF problem. The proposed algorithm reduces the size of the OPF problem and its computation costs. The simulation studies show that the proposed algorithm can efficiently and quickly separate active and inactive bus voltage and branch flow constraints solely based on reading the predicted nodal real and reactive power demand. The results show that more than 99% of voltage and branch constraints are predicted correctly and omitting them results in a significant time-saving for solving AC OPF. Further analysis of the very small fraction (less than 1%) of misclassified constraints shows that these constraints are not heavily binding and their corresponding impact on the OPF feasible space is negligible and thus affects the OPF solution very slightly.

We have tested several learning algorithms, generated diverse samples to ensure that the learners observe various patterns in the training phase, and trained learners with different hyperparameters to obtain high-quality results with a low false-negative percentage. Another reason for the low false negative percentage is the small number of active constraints in power systems optimization problems that would help to train high-quality learners.

## VII. FUTURE WORK

Advanced approaches, such as generative adversarial networks [31-33], can be used to produce more realistic operating scenarios to form a training database. Also, other constraints such as transformers constraints, phase shifter constraints, load shedding constraints, power electronic converter constraints, capacitor banks, FACTS devices, and battery storage constraints can be included in OPF, and classifiers can be used to identify inactive constraints and drop them from the optimization formulation. In addition, the proposed algorithm can be applied to other power system scheduling problems, such as unit commitment, to reduce their computational burden.

For simulation studies, we have used the default setting for classification learners' objective functions and obtained fairly accurate results. A research direction is to investigate a penalization strategy to increase the weight of false-negative classes in learners' objective functions and reduce the possibility of misclassification of true active constraints. This would be useful for problems with a high percentage of active constraints as compared to total constraints. In this paper, we have considered a fixed grid topology and constant units' production costs. Another possible research direction is to develop combined learning techniques and system models to consider grid topology and units' production cost changes in active/inactive constraints prediction. This direction is suitable for the application of the proposed algorithm on electricity market problems. In addition, to enhance the solution speed for DC OPF, one can investigate identifying the status of all inequality constraints and then solving the first-order optimality conditions based on the system of linear equations instead of solving a truncated DC OPF using optimization techniques.